\documentclass{aa}
\usepackage{graphicx}
\usepackage{txfonts}
\usepackage{amsmath}
\usepackage{gensymb}
\usepackage{hyperref}
\usepackage{color}  %SdM: allows fancy color names

\def\msun{{\rm ~M}_{\odot}}

\def\kms{{\rm ~km} {\rm ~s}^{-1}}

\def\mone{{\rm M}_1}
\def\mtwo{{\rm M}_2}
\def\mf{{\rm M}_{\rm f}}
\def\aone{{\rm a}_1}
\def\atwo{{\rm a}_2}
\def\vk{{\rm v}_{\rm k}}
\def\af{{\rm a}_{\rm f}}
\def\thf{\theta_{\rm f}}
\def\orbl{{\rm L}}
\def\prob{{\rm P}}
\def\afcrit{{\rm a}_{\rm f,crit}}
\def\vperpm{{\rm v}_{\perp m}}
\def\vperps{{\rm v}_{\perp s}}
\def\vpar{{\rm v}_\|}
\def\afpar{{\rm a}_{\rm f}^\|}
\newcommand{\eg}{{\it e.g.}}

\newcommand{\ie}{{\it i.e.}}

\newcommand{\beq}{\begin{equation}}
\newcommand{\eeq}{\end{equation}}
\newcommand{\balg}{\begin{align}}
\newcommand{\ealg}{\end{align}}
\begin{document}

\title{Formation of low-spinning $100\msun$ black holes}
\titlerunning{On the formation of low-spinning $100\msun$ black holes}
\author{
       K. Belczynski\inst{1}\thanks{Corresponding author. E-mail: chrisbelczynski@gmail.com}
       \and
       S. Banerjee\inst{2,3}
       }
\institute{
        Nicolaus Copernicus Astronomical Centre of the Polish Academy of Sciences,
        ul. Bartycka 18,
        00-716 Warszawa, Poland
        \and
        Helmholtz-Instituts f\"ur Strahlen- und Kernphysik,
        Nussallee 14-16, D-53115 Bonn, Germany
        \and
        Argelander-Institut f\"ur Astronomie,
        Auf dem H\"ugel 71, D-53121, Bonn, Germany
          }

\abstract
{}
{It is speculated that a merger of two massive stellar-origin black holes in 
a dense stellar environment may lead to the formation of a massive black hole 
in the pair-instability mass gap ($\sim 50-135\msun$). Such a merger-formed 
black hole is expected to typically have a high spin ($a \sim 0.7$). If such 
a massive black hole acquires another black hole it may lead to another merger 
detectable by LIGO/Virgo in gravitational waves. 
Acquiring a companion may be hindered by gravitational-wave kick/recoil,
which accompanies the first merger and may quickly remove the massive black 
hole from its parent globular or nuclear cluster. We test whether it is possible 
for a massive merger-formed black hole in the pair-instability gap to be retained 
in its parent cluster and have low spin. Such a black hole would be indistinguishable 
from a primordial black hole.}
{We employed results from numerical relativity calculations of black hole 
mergers to explore the range of gravitational-wave recoil velocities for various 
combinations of merging black hole masses and spins. We compared merger-formed 
massive black hole speeds with typical escape velocities from globular and
nuclear clusters.}
{We show that a  globular cluster is highly unlikely to   form and retain a $\sim 100\msun$ black 
hole if the spin of the black hole is low ($a \lesssim 0.3$). 
Massive merger-formed black holes with low spins acquire high recoil speeds 
($\gtrsim 200\kms$) from gravitational-wave kick during formation that exceed 
typical escape speeds from globular clusters ($\sim 50\kms$).   
However, a very low-spinning ($a \sim 0.1$) and massive ($\sim 100\msun$) black 
hole could be formed and retained in a galactic nuclear  star cluster. Even though 
such massive merger-formed black holes with such low spins acquire high 
speeds during formation ($\sim 400 \kms$), they may avoid ejection since massive 
nuclear clusters have high escape velocities ($\sim 300-500\kms$).
A future detection of a massive black hole in the pair-instability mass gap with 
low spin would therefore not be proof of the existence of primordial black holes, which 
are sometimes claimed to have low spins and arbitrarily high masses.}
{}


\maketitle

\section{Introduction}\label{sec.intro}

By the beginning of 2020 the LIGO/Virgo collaboration had published  ten black
hole--black hole (BH--BH) mergers from the first two (O1/O2) observational 
runs~\citep{LIGO2019a}, and more than $50$ BH--BH merger alerts from their recently 
concluded third observing run (O3; see~\url{https://gracedb.ligo.org/superevents/public/O3/}).
The origin of these BH--BH mergers is still unknown.

Numerous formation scenarios and sites have been proposed in the literature to explain 
LIGO/Virgo BH--BH merger detections. The two most prominent proposals are the classical
isolated binary evolution channel in galactic fields~\citep{Tutukov1993,Lipunov1997,
Voss2003} and dynamical formation in dense clusters~\citep{Miller2002a,
PortegiesZwart2004,Gultekin2006}. Triple stars are also proposed as BH--BH merger 
formation sites~\citep{Antonini2018}, or isolated binaries with very rapid rotation 
leading to homogeneous evolution~\citep{Marchant2016}. First generation of stars is 
yet another potential formation site~\cite{Bond1984b}.

Massive stars are subject to pair instabilities~\citep{Bond1984a} and are not expected 
to form BHs with masses of $\sim 100\msun$. We note that typically this pair-instability 
mass gap is adopted to be broad, namely $\sim 50-135\msun$, but recently it was claimed that 
the range may be narrower: $\sim 90-125\msun$~\citep{Farmer2020}. The stellar-origin 
BHs are claimed to form with low spins $a \lesssim 0.2$ from single massive stars
~\citep{Fuller2019a,Belczynski2020b}. Mergers of two massive BHs in dense clusters can 
produce $\sim 100\msun$ BHs, but such BHs will typically have large spin $a \sim 0.7$
~\citep{Gerosa2017,Fishbach2017a}. Therefore, it seems that astrophysical channels that 
invoke stars, at least initially, to produce BHs cannot form BHs of  $\sim 100\msun$
 with 
low spins. This inference could be potentially used to distinguish stellar-origin BHs 
from primordial BHs.  

Primordial BHs were proposed to form from density fluctuations in the very early Universe 
by~\cite{Hawking1971}. These primordial BHs are sometimes employed to explain LIGO/Virgo 
BH--BH mergers, that is if they catch a companion BH \citep{Clesse2017}. Although in principle 
primordial BHs could have an arbitrary mass, various observational constraints exclude 
most possibilities, nevertheless allowing masses in the range $\sim 10-100\msun$~\citep{Carr2018}. 
Primordial BHs are usually argued to have very low spins $a \sim 0.01$~\citep{DeLuca2019}. 

It may appear that the detection of a $100\msun$ BH with low spin (e.g., by LIGO/Virgo) 
would point to the primordial formation scenario. However, there are several alternative 
astrophysical scenarios. 
It has been proposed in the past that numerous consecutive mergers of a seed BH with 
lighter companions (either other BHs or stars) with randomly oriented spins (so 
they cancel out in the final merger product) would not increase the spin of the 
seed BH~\citep{Miller2002,Hughes2003,Antonini_2019}. 
Consecutive (runaway) mergers of stars in young stellar clusters were claimed to 
lead to the formation of quasi-stellar objects massive enough to form $\sim 100\msun$ 
BHs~\citep{Portegies_2007,Fujii_2013,DiCarlo2019}. However, it is not clear how 
such a quasi-stellar object would avoid pair-instability explosion (no BH formation) 
or what  the spin would be of a BH if pair instability was avoided somehow.

Here we investigate the validity of a different astrophysical scenario. We test the 
possibility that a merger of two massive stellar-origin BHs can produce a $\sim 100\msun$ 
BH with low spin and with such spatial velocity that it could be retained in its parent 
globular or nuclear cluster. Such a BH could catch another BH and produce a merger 
detectable by  LIGO/Virgo.

\section{Calculations}\label{sec.calc}

We explore a range of mass ratios of BH--BH mergers, leading to a BH of 
$\approx100\msun$, by considering three combinations of component masses, 
namely $(\mone,\mtwo)=(50.0\msun,50.0\msun)$, $(70.0\msun,35.0\msun)$, and 
$(70.0\msun,20.0\msun)$. These choices are motivated by the range of mass ratios in 
BH--BH mergers observed to date ($\approx0.3{\rm ~-}\approx 1.0$ \cite{LIGO2019b}).
For each combination, we consider two extreme cases, namely both component BHs are 
maximally spinning, \ie, their dimensionless spin magnitudes are 
$(\aone,\atwo)=(1.0,1.0)$ or both components are not spinning, \ie, 
$(\aone,\atwo)=(0.0,0.0)$. For each $(\mone,\mtwo),(\aone,\atwo)$ combination,
we perform $10^5$ evaluations of the final merged BH's dimensionless spin vector, 
$\vec\af$, and the recoil kick velocity, $\vec\vk$, due to the anisotropic radiation 
of GW during the merger assuming that the component BH spins are uncorrelated and 
oriented isotropically with respect to the orbital angular momentum. We note that 
the energy carried away by the GW emission will result in the final BHs' mass, 
$\mf$, being smaller than $\mone+\mtwo$. Since this discrepancy is at most a few 
percent (but depends, in a complex manner, on the BH-BH binary mass ratios and 
spin vectors as numerical-relativity calculations suggest \cite{Hughes_2009,Sperhake_2015}), 
we ignore it for the present purpose and assign $\mf\approx\mone+\mtwo$.

The vectors $\vec\vk$ and $\vec\af$ of the merged BH were obtained using 
numerical relativity (NR) calculations for a wide range of mass ratio 
and spin configurations of the merging BHs \citep{Pretorius_2005,Campanelli_2007,
Baker_2008,vanMeter_2010} and fitting formulae exist that well reproduce the NR 
outcomes. In general, if the spins of the merging BHs  are zero, $\vec\vk$ will lie on 
the orbital plane and be aligned along the line joining the BHs just before the 
merger. Its magnitude, $\vk$, is zero (small) for equal-mass (extreme-mass-ratio) 
components and maximizes to $\approx170\kms$ at the mass ratio of $\approx1/2.9$. If the 
merging BHs have spins, then depending on the  magnitudes and orientations of the spins, 
$\vec\vk$ will also have an in-plane component perpendicular to the mass axis 
and a component perpendicular to the orbital plane. For (near-) maximally spinning
BHs, this off-plane component of the merger recoil typically dominates and can well 
exceed $500\kms$ \citep{Baker_2008,vanMeter_2010}; for certain configurations, it 
can reach $\approx3000\kms$ \citep{Campanelli_2007}. In this work, we use the 
NR-based fitting formulae of \citet{vanMeter_2010} for the components of $\vec\vk$, which 
incorporate cases where the  spins of the BHs are inclined with respect to the orbital angular 
momentum and would therefore undergo spin-orbit precession during the in-spiral 
and merger phases of the BH--BH. These formulae agree with NR outcomes within $5\%$.

If the spins of the merging BH components are zero, then the only source of angular 
momentum of the merged BH is the BH-BH orbital angular momentum. Accordingly, the
dimensionless spin of the merged BH, $\vec\af$, will be aligned with the orbital
angular momentum, $\vec\orbl$, and, for equal-mass components, will have the magnitude
$\af\approx0.7$ \citep{Pretorius_2005}. When the merging BHs have finite, misaligned 
spins, $\vec\af$ is generally misaligned relative to $\vec\orbl$ and its magnitude is 
augmented (suppressed) with respect to the nonspinning-merger value if the spins are 
pro-aligned (anti-aligned). For most configurations, $\vec\orbl$ dominates the angular 
momentum budget of the BH--BH system \citep{Rezzolla_2008} and hence the angle between 
$\vec\af$ and $\vec\orbl$ is typically $<10\degree$. In this work, we use the
NR-based fitting functions of \citet{Rezzolla_2008} to compute $\vec\af$ for each of 
the randomly chosen component-spin orientations. The NR-based fitting formulae
used in this work are described in Sect.~\ref{sec.nrform}.

For each combination, $(\mone,\mtwo),(\aone,\atwo)$, of the merging BH masses
and dimensionless spin magnitudes (see above), we perform $10^5$ evaluations of $\vec\af$
and $\vec\vk$. Each evaluation proceeds as follows: the inclinations and the
azimuths of the component BH spins, $(\theta_1,\theta_2)$, $(\phi_1,\phi_2)$,
with respect to the orbital angular momentum, $\vec\orbl$,
are chosen randomly and independently of each other from a uniform distribution
within the range $(0,2\pi)$. Next, the NR fitting formulae, as described in Sect.~\ref{sec.nrform},
are applied to
evaluate the in-plane and off-plane components of the vector, $\vec\vk$,
of the merger recoil kick and the magnitude of the final spin, $\af$.
We used dedicated numerical routines for this purpose,
in which the formulae in Sect.~\ref{sec.nrform} are implemented
to evaluate the $\vec\vk$-components and $\af$ (and $\thf$) simultaneously.
Given the pure analytical nature and medium complexity
of these expressions, each evaluation sequence takes less than a second;
$10^5$ iterations typically take $<3$ hr 
on a single $\sim3$ Ghz processor. That way, the distributions
of $\vk$ and $\af$ for each $(\mone,\mtwo),(\aone,\atwo)$
combination are obtained for isotropic and uncorrelated orientations
of the spins of the merging BHs. As explained in Sect.~\ref{sec.nrform},
the effect of the undetermined angles on the fitting formulae
is automatically cancelled out by the choice of a large number of
isotropically distributed spin orientations. Therefore, the distributions
of $\vk$ and $\af$ depend only on the $(\mone,\mtwo),(\aone,\atwo)$ combination.

\begin{figure}
\hspace*{-0.4cm}
\includegraphics[width=0.5\textwidth]{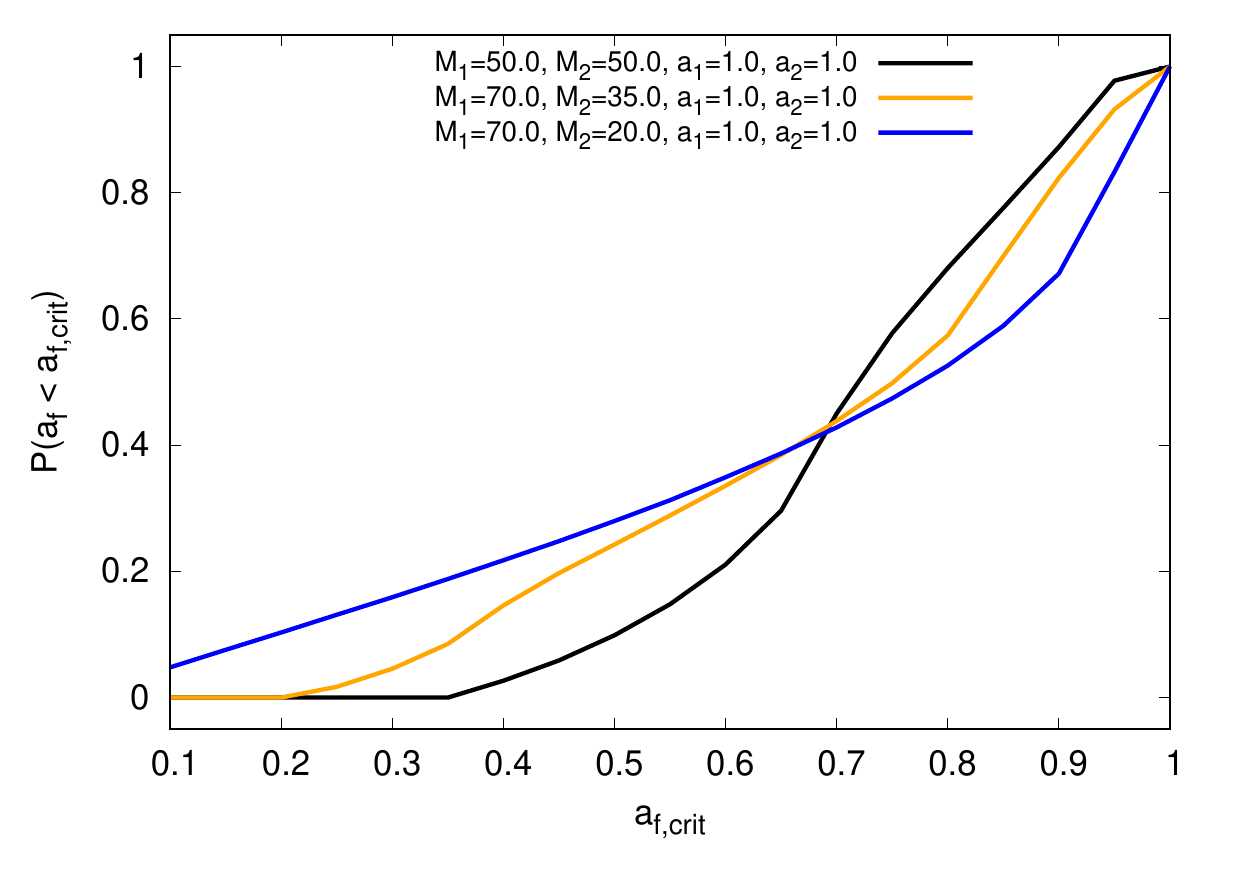}
\caption{ 
Probability, $\prob(\af<\afcrit)$, that the dimensionless spin magnitude,
$\af$, of a merged BH is less than the value $\afcrit$, as a function of 
$\afcrit$. Each curve corresponds to a particular mass combination of the 
maximally spinning components of the merging BH--BH, as indicated in the legend. 
The probabilities are obtained out of $10^5$ trials of random and uncorrelated 
orientations of the spins of the BH components.
}
\label{fig.1}
\end{figure}

Figure~\ref{fig.1} shows the probability, ${\rm P}$, that $\af<\afcrit$ as a function
of $\afcrit$, with $\afcrit$ being a chosen threshold value of $\af$, for the 
three component-mass combinations and for $\aone=\atwo=1.0$. The fractions plotted 
in this figure are out of $10^5$ trials (see above). Because of the dominance of 
$\vec\orbl$, the final angular momentum of the BH--BH system can be significantly 
suppressed only for strongly anti-aligned spin orientations of the merging BHs.
Accordingly, the fraction of configurations that suppress the system's angular 
momentum to produce $\af$ below a certain $\afcrit$ decreases strongly with 
$\afcrit$, as Fig.~\ref{fig.1} demonstrates. Only mergers with the most dissimilar 
BH masses $(70.0\msun,20.0\msun)$ can yield $\af<0.1$, and that too with a very small 
probability ($\prob\approx0.05$). Equal-mass, maximally spinning BHs cannot yield 
$\af$ below $\approx 0.35$ and, with the mass ratio of $0.5$, the lower limit of 
$\af$ is $\approx0.2$.

\begin{figure}
\hspace*{-0.4cm}
\includegraphics[width=0.5\textwidth]{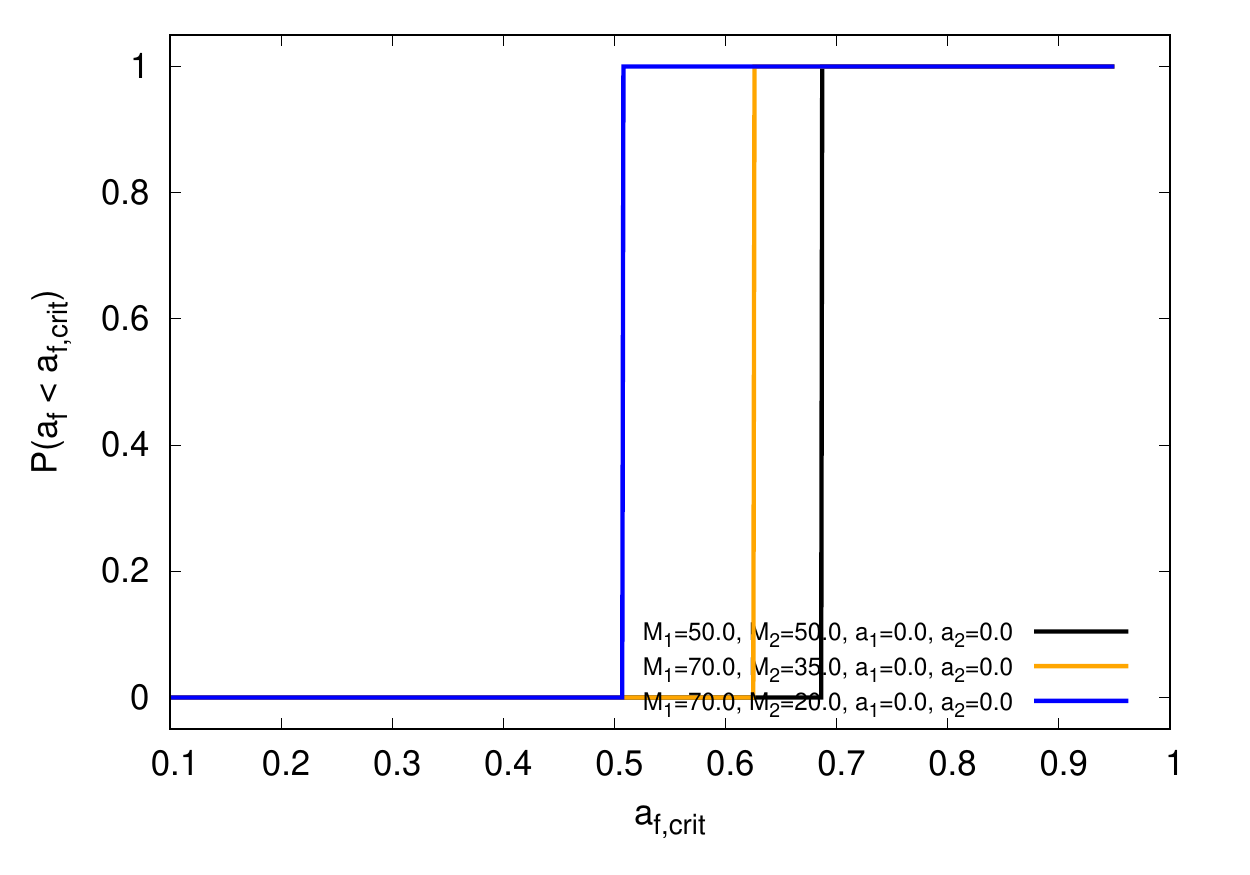}
\caption{ 
Same plot as Fig.~\ref{fig.1} but for nonspinning BH components.
}
\label{fig.2}
\end{figure}

On the other hand, when $\aone=\atwo=0.0$ the orbital angular momentum,
which is necessarily perpendicular to the orbital plane, is the
only resource of angular momentum of the merging BH--BH system. In that case,
both $\af$ and $\vk$ are unique and depend only on the mass ratio of  
the merging BHs, as determined by the NR fitting formulae employed here.
Therefore, $\prob(\af<\afcrit)$ is a step function jumping from zero to unity 
at the value of $\afcrit$ that is equal to the unique $\af$ for the chosen mass 
ratio. This is demonstrated in Fig.~\ref{fig.2}. With nonspinning BHs, the lowest 
$\af$ obtained here is $\approx0.5$ for mergers with the smallest mass 
ratio ($\approx0.3$).

\begin{figure}
\hspace*{-0.4cm}
\includegraphics[width=0.5\textwidth]{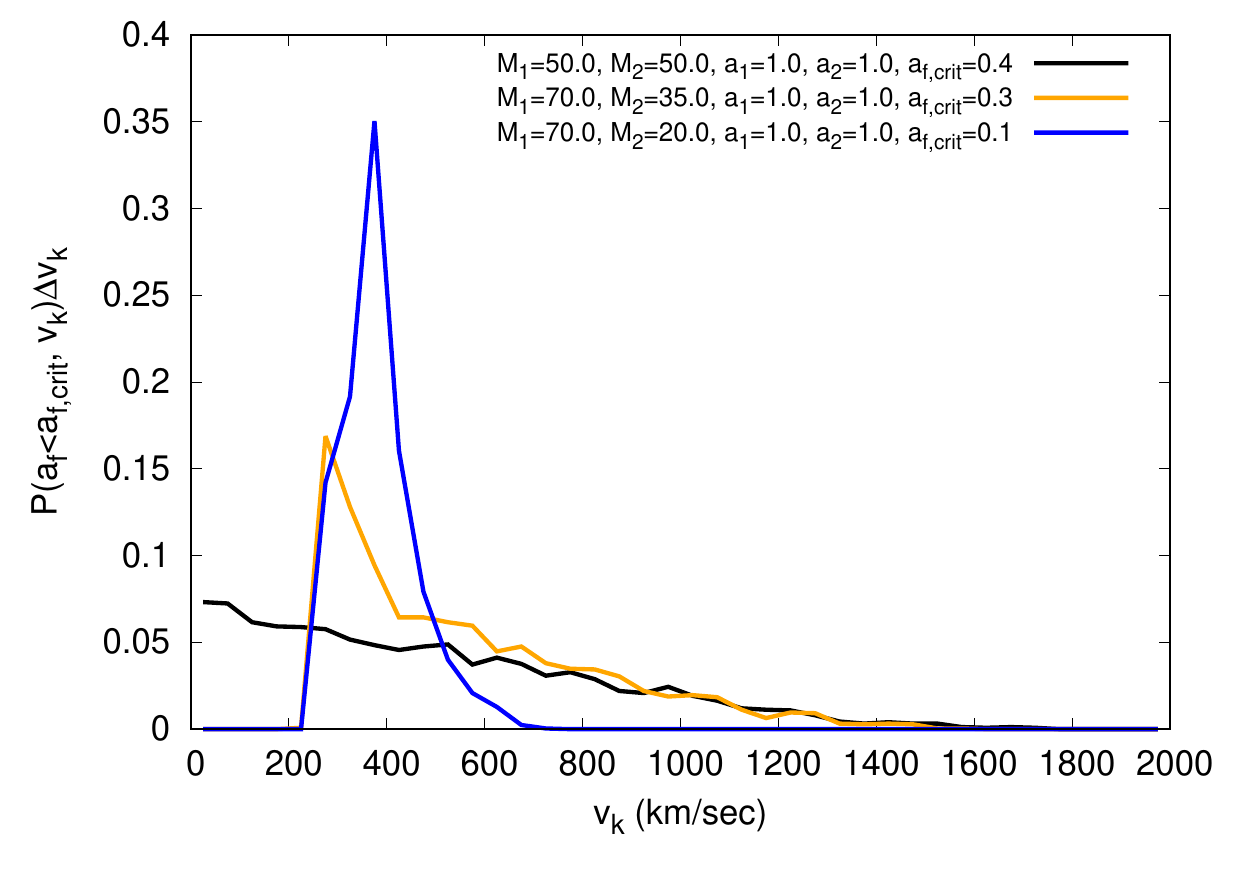}
\caption{ 
Distributions of GW recoil kick magnitudes, $\vk$, for those mergers
of maximally spinning BHs that yield a low-spinning merged BH with 
$\af<\afcrit$. The BH-mass combinations and the correspondingly chosen 
$\afcrit$s are indicated in the legend.
}
\label{fig.3}
\end{figure}

Since our focus here is the low-spinning BH--BH merger products, it would be worth 
considering the GW recoil kick magnitudes, $\vk$, such merged BHs would receive as 
a result of the merger. This is shown in Fig.~\ref{fig.3}. Here, the distributions 
of $\vk$ are shown for those $\aone=\atwo=1.0$ mergers that lead to $\af<0.1$, 
$0.3$, and $0.4$ for the mass ratios $\approx0.3$, $0.5$, and $1.0,$ respectively 
(for $\aone=\atwo=0.0$, we have $\af>0.5$; see above). As evident, the strongly 
anti-aligned mergers that yield $\af<0.1$ have a relatively narrow range of recoil 
kicks, $200\kms\lesssim\vk\lesssim700\kms$, and their distribution is strongly 
peaked at $\vk\approx400\kms$. For $\afcrit=0.3$ and $0.4$ and mass ratios 
$\gtrsim 0.5$, the $\vk$ distributions are peaked at smaller values but have tails 
extending up to $1600\kms$. Qualitatively, any BH--BH configuration (out of spinning 
BH components) that leads to a small $\af$ would emit GW relatively symmetrically 
leading to a relatively low recoil. But due to the required low mass ratio (see 
Fig.~\ref{fig.1}), the configuration just before the merger is still asymmetrical 
leading to recoils of typically 400$\kms$.

\section{Discussion and Conclusions}\label{sec:concl} 

We have shown that a merger of two rapidly spinning ($a_1 \approx a_2 \approx 1$) 
and unequal mass ($q \lesssim 0.3$) BHs can produce a $\sim 100\msun$ BH with low 
spin and sufficiently low spatial velocity (from GW kick) for it to be 
retained in a nuclear cluster. Such a BH, under the assumption that it catches 
another BH, can manifest itself in high-frequency GW observations (LIGO/Virgo).  

So far, (O1/O2) LIGO/Virgo observations seem to indicate that the detected BH--BH 
mergers have component BHs with low spins (\citealt{LIGO2019b}; see also 
\citealt{Belczynski2020b,Banerjee_2020b}), with the exception of GW170729 and GW190412 
which are consistent with a moderately spinning BH~\citep{LIGO2019b,gw190412}. This 
was explained in terms of Tayler-Spruit magnetic dynamo~\citep{Spruit2002,Fuller2019a} 
which efficiently removes angular momentum from massive stars and leads to the formation 
of low-spinning BHs in BH--BH mergers~\citep{Belczynski2020b,Bavera2020}. If this applies 
to all first-generation BHs, then the only way to produce a merger of two highly spinning 
BHs is to have a merger of two second-generation BHs, as these will typically have large 
spin~\citep{Gerosa2017,Fishbach2017a}.
Alternatively, it seems that nature can produce rapidly spinning BHs, as these are 
observed in Galactic and extra-galactic high-mass X-ray binaries (HMXRBs): Cyg 
X-1: $a=0.98$, LMC X-1: $a=0.92$, and M33 X-7: $a=0.84$~\citep{Miller2015b}. 
The apparent tension between LIGO/Virgo BHs and those in HMXRBs can be understood 
if the specific evolutionary scenarios are different for both populations
(e.g., ~\cite{Qin2019}). Whether or not the tension is 
fully understood (which is still debated), it is clear that nature can
produce both slow- and rapidly spinning BHs, and if the rapidly spinning BHs 
merge they are the potential candidates for our scenario. 

Stars can possibly form BHs massive enough to satisfy the total mass requirement 
and the uneven mass ratio limit. As noted in Sect.~\ref{sec.intro}, stars can 
possibly form BHs (first generation) as massive as $\sim 90\msun$ and still 
avoid pair-instability pulsation mass loss. Such a massive BH would then need to 
dynamically pair with a $\sim 20-30\msun$ BH in a dense stellar cluster,
which is possible as Monte Carlo and direct N-body computations show 
\citep{Rodriguez2019b,DiCarlo2019,Banerjee_2020b}. As we 
demonstrate here, the formation of a low-spinning $\sim 100\msun$ BH induces a strong 
GW recoil kick that accelerates the merged BH to a high speed of $\sim 400\kms$. 
This removes all such BHs from globular clusters
(mass $\sim 10^5-10^6\msun$, size $\sim$ pc; \citealt{Harris1996,Baumgardt_2018}), which have typical central escape velocities of $\lesssim 70 \kms$ \citep[\eg,][]{Georgiev_2009}.
However, such BHs can 
potentially be retained in nuclear clusters (mass $\sim 10^6-10^8\msun$, size $\sim$ pc;
\citealt{Schoedel_2014})
that typically have much higher central escape velocities of 
$\sim 300-500\kms$~\citep{Georgiev2016,Schoedel_2014,Georgiev_2009,Miller2009}.

We note that current LIGO/Virgo observations do not provide spin magnitude 
measurements of merging BHs. These observations provide measurements of the BH--BH 
effective spin parameter, which is a projection of both BH spins weighted by 
BH mass on the binary's orbital angular momentum vector. This means that if the 
effective spin is low then the BH spins are low; the spin vectors 
are facing in opposite directions causing them to cancel out; or the spin 
vectors are in the orbital plane. However, future observations may provide 
large numbers of massive BHs and this may allow for statistical assessment of 
BH spin magnitudes. Alternatively, there is hope for measuring spin precession 
in the inspiral waveform in future. This would also allow the spin 
magnitudes of merging BHs to be estimated.

In future work, we intend to explore the effect of more recent NR results (\eg, 
\cite{Lousto_2013, Hofmann_2016})  and the mass--spin dependence of BHs as 
obtained from stellar-evolutionary calculations \citep{Belczynski2020b}.
Also, the formation rate estimate of the proposed scenario would require careful 
analysis of nuclear cluster evolution folded with detailed stellar and binary 
evolution. Such an estimate is beyond the scope of this paper.

Here, we simply point out (proof-of-principle) the possibility of the formation of 
a $100\msun$, low-spinning BH of astrophysical origin. This adds to existing 
scenarios of a seed BH growing in mass by accretion of lower mass stars or light 
BHs~\citep{Miller2002,Hughes2003} and BH formation from a runaway stellar merger 
product \citep{Portegies_2007,Fujii_2013,DiCarlo2019}.
These astrophysical scenarios are to be contrasted with any claims that such BHs 
(if they exist and are detected) must be of primordial origin. Despite the 
fact that the channel pointed out here is potentially of lower probability than 
the other channels, it follows from basic physics (numerical relativity) and 
requires only a specific astrophysical site (nuclear cluster), of which there are 
many in the Universe. 

\begin{acknowledgements}
We thank the anonymous referee for constructive comments and useful
suggestions that have helped to improve the manuscript. 
KB acknowledges support from the Polish National Science Center (NCN) grant
Maestro (2018/30/A/ST9/00050).
SB acknowledges the support from the Deutsche Forschungsgemeinschaft (DFG;
German Research Foundation) through the individual research grant ``The
dynamics of stellar-mass black holes in dense stellar systems and their
role in gravitational-wave generation'' (BA 4281/6-1; PI: S. Banerjee).
The authors thank Cole Miller and Mirek Giersz for useful comments
and criticisms that have helped to improve the manuscript. They
acknowledge the hospitality of the National Astronomical Observatories
of China (NAOC), Beijing where a good part of this work has been done.
\end{acknowledgements}

\bibliographystyle{aa}
\bibliography{biblio.bib}

\begin{thebibliography}{56}
\expandafter\ifx\csname natexlab\endcsname\relax\def\natexlab#1{#1}\fi

\bibitem[{{Abbott} {et~al.}(2019{\natexlab{a}}){Abbott}, {Abbott}, {Abbott},
  {Abraham}, {LIGO Scientific Collaboration}, \& {Virgo
  Collaboration}}]{LIGO2019a}
{Abbott}, B.~P., {Abbott}, R., {Abbott}, T.~D., {et~al.} 2019{\natexlab{a}},
  \apjl, 882, L24

\bibitem[{{Abbott} {et~al.}(2019{\natexlab{b}}){Abbott}, {Abbott}, {Abbott},
  {Abraham}, {LIGO Scientific Collaboration}, \& {Virgo
  Collaboration}}]{LIGO2019b}
{Abbott}, B.~P., {Abbott}, R., {Abbott}, T.~D., {et~al.} 2019{\natexlab{b}},
  Physical Review X, 9, 031040

\bibitem[{{Antonini} {et~al.}(2019){Antonini}, {Gieles}, \&
  {Gualandris}}]{Antonini_2019}
{Antonini}, F., {Gieles}, M., \& {Gualandris}, A. 2019, \mnras, 486, 5008

\bibitem[{{Antonini} {et~al.}(2018){Antonini}, {Rodriguez}, {Petrovich}, \&
  {Fischer}}]{Antonini2018}
{Antonini}, F., {Rodriguez}, C.~L., {Petrovich}, C., \& {Fischer}, C.~L. 2018,
  \mnras, 480, L58

\bibitem[{{Baker} {et~al.}(2007){Baker}, {Boggs}, {Centrella}, {Kelly},
  {McWilliams}, {Miller}, \& {van Meter}}]{Baker_2007}
{Baker}, J.~G., {Boggs}, W.~D., {Centrella}, J., {et~al.} 2007, \apj, 668, 1140

\bibitem[{{Baker} {et~al.}(2008){Baker}, {Boggs}, {Centrella}, {Kelly},
  {McWilliams}, {Miller}, \& {van Meter}}]{Baker_2008}
{Baker}, J.~G., {Boggs}, W.~D., {Centrella}, J., {et~al.} 2008, \apjl, 682, L29

\bibitem[{{Banerjee}(2020)}]{Banerjee_2020b}
{Banerjee}, S. 2020, arXiv e-prints, arXiv:2004.07382

\bibitem[{{Baumgardt} \& {Hilker}(2018)}]{Baumgardt_2018}
{Baumgardt}, H. \& {Hilker}, M. 2018, \mnras, 478, 1520

\bibitem[{{Bavera} {et~al.}(2020){Bavera}, {Fragos}, {Qin}, {Zapartas},
  {Neijssel}, {Mandel}, {Batta}, {Gaebel}, {Kimball}, \&
  {Stevenson}}]{Bavera2020}
{Bavera}, S.~S., {Fragos}, T., {Qin}, Y., {et~al.} 2020, \aap, 635, A97

\bibitem[{{Belczynski} {et~al.}(2020){Belczynski}, {Klencki}, {Fields},
  {Olejak}, {Berti}, {Meynet}, {Fryer}, {Holz}, {O'Shaughnessy}, {Brown},
  {Bulik}, {Leung}, {Nomoto}, {Madau}, {Hirschi}, {Kaiser}, {Jones}, {Mondal},
  {Chruslinska}, {Drozda}, {Gerosa}, {Doctor}, {Giersz}, {Ekstrom}, {Georgy},
  {Askar}, {Baibhav}, {Wysocki}, {Natan}, {Farr}, {Wiktorowicz}, {Coleman
  Miller}, {Farr}, \& {Lasota}}]{Belczynski2020b}
{Belczynski}, K., {Klencki}, J., {Fields}, C.~E., {et~al.} 2020, \aap, 636,
  A104

\bibitem[{{Blanchet} {et~al.}(2005){Blanchet}, {Qusailah}, \&
  {Will}}]{Blanchet_2005}
{Blanchet}, L., {Qusailah}, M. S.~S., \& {Will}, C.~M. 2005, \apj, 635, 508

\bibitem[{{Bond} {et~al.}(1984){Bond}, {Arnett}, \& {Carr}}]{Bond1984a}
{Bond}, J.~R., {Arnett}, W.~D., \& {Carr}, B.~J. 1984, \apj, 280, 825

\bibitem[{{Bond} \& {Carr}(1984)}]{Bond1984b}
{Bond}, J.~R. \& {Carr}, B.~J. 1984, \mnras, 207, 585

\bibitem[{{Campanelli} {et~al.}(2007){Campanelli}, {Lousto}, {Zlochower}, \&
  {Merritt}}]{Campanelli_2007}
{Campanelli}, M., {Lousto}, C., {Zlochower}, Y., \& {Merritt}, D. 2007, \apjl,
  659, L5

\bibitem[{{Carr} \& {Silk}(2018)}]{Carr2018}
{Carr}, B. \& {Silk}, J. 2018, \mnras, 478, 3756

\bibitem[{{Clesse} \& {Garc{\'\i}a-Bellido}(2017)}]{Clesse2017}
{Clesse}, S. \& {Garc{\'\i}a-Bellido}, J. 2017, Physics of the Dark Universe,
  15, 142

\bibitem[{{De Luca} {et~al.}(2019){De Luca}, {Desjacques}, {Franciolini},
  {Malhotra}, \& {Riotto}}]{DeLuca2019}
{De Luca}, V., {Desjacques}, V., {Franciolini}, G., {Malhotra}, A., \&
  {Riotto}, A. 2019, Journal of Cosmology and Astroparticle Physics, 2019, 018

\bibitem[{{Di Carlo} {et~al.}(2019){Di Carlo}, {Giacobbo}, {Mapelli},
  {Pasquato}, {Spera}, {Wang}, \& {Haardt}}]{DiCarlo2019}
{Di Carlo}, U.~N., {Giacobbo}, N., {Mapelli}, M., {et~al.} 2019, \mnras, 487,
  2947

\bibitem[{{Farmer} {et~al.}(2020){Farmer}, {Renzo}, {de Mink}, {Fishbach}, \&
  {Justham}}]{Farmer2020}
{Farmer}, R., {Renzo}, M., {de Mink}, S., {Fishbach}, M., \& {Justham}, S.
  2020, arXiv e-prints, arXiv:2006.06678

\bibitem[{{Favata} {et~al.}(2004){Favata}, {Hughes}, \& {Holz}}]{Favata_2004}
{Favata}, M., {Hughes}, S.~A., \& {Holz}, D.~E. 2004, \apjl, 607, L5

\bibitem[{{Fishbach} {et~al.}(2017){Fishbach}, {Holz}, \&
  {Farr}}]{Fishbach2017a}
{Fishbach}, M., {Holz}, D.~E., \& {Farr}, B. 2017, \apjl, 840, L24

\bibitem[{{Fitchett}(1983)}]{Fitchett_1983}
{Fitchett}, M.~J. 1983, \mnras, 203, 1049

\bibitem[{{Fujii} \& {Portegies Zwart}(2013)}]{Fujii_2013}
{Fujii}, M.~S. \& {Portegies Zwart}, S. 2013, \mnras, 430, 1018

\bibitem[{{Fuller} {et~al.}(2019){Fuller}, {Piro}, \& {Jermyn}}]{Fuller2019a}
{Fuller}, J., {Piro}, A.~L., \& {Jermyn}, A.~S. 2019, \mnras, 485, 3661

\bibitem[{{Georgiev} {et~al.}(2016){Georgiev}, {B{\"o}ker}, {Leigh},
  {L{\"u}tzgendorf}, \& {Neumayer}}]{Georgiev2016}
{Georgiev}, I.~Y., {B{\"o}ker}, T., {Leigh}, N., {L{\"u}tzgendorf}, N., \&
  {Neumayer}, N. 2016, \mnras, 457, 2122

\bibitem[{{Georgiev} {et~al.}(2009){Georgiev}, {Hilker}, {Puzia}, {Goudfrooij},
  \& {Baumgardt}}]{Georgiev_2009}
{Georgiev}, I.~Y., {Hilker}, M., {Puzia}, T.~H., {Goudfrooij}, P., \&
  {Baumgardt}, H. 2009, \mnras, 396, 1075

\bibitem[{Gerosa \& Berti(2017)}]{Gerosa2017}
Gerosa, D. \& Berti, E. 2017, Phys. Rev., D95, 124046

\bibitem[{{Gonz{\'a}lez} {et~al.}(2007){Gonz{\'a}lez}, {Sperhake},
  {Br{\"u}gmann}, {Hannam}, \& {Husa}}]{Gonzalez_2007}
{Gonz{\'a}lez}, J.~A., {Sperhake}, U., {Br{\"u}gmann}, B., {Hannam}, M., \&
  {Husa}, S. 2007, \prl, 98, 091101

\bibitem[{{G{\"u}ltekin} {et~al.}(2006){G{\"u}ltekin}, {Miller}, \&
  {Hamilton}}]{Gultekin2006}
{G{\"u}ltekin}, K., {Miller}, M.~C., \& {Hamilton}, D.~P. 2006, \apj, 640, 156

\bibitem[{{Harris}(1996)}]{Harris1996}
{Harris}, W.~E. 1996, \aj, 112, 1487

\bibitem[{{Hawking}(1971)}]{Hawking1971}
{Hawking}, S. 1971, \mnras, 152, 75

\bibitem[{{Hofmann} {et~al.}(2016){Hofmann}, {Barausse}, \&
  {Rezzolla}}]{Hofmann_2016}
{Hofmann}, F., {Barausse}, E., \& {Rezzolla}, L. 2016, \apjl, 825, L19

\bibitem[{{Hughes}(2009)}]{Hughes_2009}
{Hughes}, S.~A. 2009, \araa, 47, 107

\bibitem[{{Hughes} \& {Blandford}(2003)}]{Hughes2003}
{Hughes}, S.~A. \& {Blandford}, R.~D. 2003, \apjl, 585, L101

\bibitem[{{Kidder}(1995)}]{Kidder_1995}
{Kidder}, L.~E. 1995, \prd, 52, 821

\bibitem[{{Lipunov} {et~al.}(1997){Lipunov}, {Postnov}, \&
  {Prokhorov}}]{Lipunov1997}
{Lipunov}, V.~M., {Postnov}, K.~A., \& {Prokhorov}, M.~E. 1997, Astronomy
  Letters, 23, 492

\bibitem[{{Lousto} \& {Zlochower}(2013)}]{Lousto_2013}
{Lousto}, C.~O. \& {Zlochower}, Y. 2013, \prd, 87, 084027

\bibitem[{{Marchant} {et~al.}(2016){Marchant}, {Langer}, {Podsiadlowski},
  {Tauris}, \& {Moriya}}]{Marchant2016}
{Marchant}, P., {Langer}, N., {Podsiadlowski}, P., {Tauris}, T.~M., \&
  {Moriya}, T.~J. 2016, \aap, 588, A50

\bibitem[{{Miller}(2002)}]{Miller2002}
{Miller}, M.~C. 2002, \apj, 581, 438

\bibitem[{{Miller} \& {Hamilton}(2002)}]{Miller2002a}
{Miller}, M.~C. \& {Hamilton}, D.~P. 2002, \mnras, 330, 232

\bibitem[{{Miller} \& {Lauburg}(2009)}]{Miller2009}
{Miller}, M.~C. \& {Lauburg}, V.~M. 2009, \apj, 692, 917

\bibitem[{{Miller} \& {Miller}(2015)}]{Miller2015b}
{Miller}, M.~C. \& {Miller}, J.~M. 2015, \physrep, 548, 1

\bibitem[{{Morawski} {et~al.}(2018){Morawski}, {Giersz}, {Askar}, \&
  {Belczynski}}]{Morawski2018}
{Morawski}, J., {Giersz}, M., {Askar}, A., \& {Belczynski}, K. 2018, \mnras,
  481, 2168

\bibitem[{{Portegies Zwart} {et~al.}(2004){Portegies Zwart}, {Baumgardt},
  {Hut}, {Makino}, \& {McMillan}}]{PortegiesZwart2004}
{Portegies Zwart}, S.~F., {Baumgardt}, H., {Hut}, P., {Makino}, J., \&
  {McMillan}, S.~L.~W. 2004, \nat, 428, 724

\bibitem[{{Portegies Zwart} \& {van den Heuvel}(2007)}]{Portegies_2007}
{Portegies Zwart}, S.~F. \& {van den Heuvel}, E. P.~J. 2007, \nat, 450, 388

\bibitem[{{Pretorius}(2005)}]{Pretorius_2005}
{Pretorius}, F. 2005, \prl, 95, 121101

\bibitem[{{Qin} {et~al.}(2019){Qin}, {Marchant}, {Fragos}, {Meynet}, \&
  {Kalogera}}]{Qin2019}
{Qin}, Y., {Marchant}, P., {Fragos}, T., {Meynet}, G., \& {Kalogera}, V. 2019,
  \apjl, 870, L18

\bibitem[{{Rezzolla} {et~al.}(2008){Rezzolla}, {Barausse}, {Dorband},
  {Pollney}, {Reisswig}, {Seiler}, \& {Husa}}]{Rezzolla_2008}
{Rezzolla}, L., {Barausse}, E., {Dorband}, E.~N., {et~al.} 2008, \prd, 78,
  044002

\bibitem[{{Rodriguez} {et~al.}(2019){Rodriguez}, {Zevin}, {Amaro-Seoane},
  {Chatterjee}, {Kremer}, {Rasio}, \& {Ye}}]{Rodriguez2019b}
{Rodriguez}, C.~L., {Zevin}, M., {Amaro-Seoane}, P., {et~al.} 2019, \prd, 100,
  043027

\bibitem[{{Sch{\"o}del} {et~al.}(2014){Sch{\"o}del}, {Feldmeier}, {Kunneriath},
  {Stolovy}, {Neumayer}, {Amaro-Seoane}, \& {Nishiyama}}]{Schoedel_2014}
{Sch{\"o}del}, R., {Feldmeier}, A., {Kunneriath}, D., {et~al.} 2014, \aap, 566,
  A47

\bibitem[{{Sperhake}(2015)}]{Sperhake_2015}
{Sperhake}, U. 2015, Classical and Quantum Gravity, 32, 124011

\bibitem[{{Spruit}(2002)}]{Spruit2002}
{Spruit}, H.~C. 2002, \aap, 381, 923

\bibitem[{{The LIGO Scientific Collaboration} \& {the Virgo
  Collaboration}(2020)}]{gw190412}
{The LIGO Scientific Collaboration} \& {the Virgo Collaboration}. 2020, arXiv
  e-prints, arXiv:2004.08342

\bibitem[{{Tutukov} \& {Yungelson}(1993)}]{Tutukov1993}
{Tutukov}, A.~V. \& {Yungelson}, L.~R. 1993, \mnras, 260, 675

\bibitem[{{van Meter} {et~al.}(2010){van Meter}, {Miller}, {Baker}, {Boggs}, \&
  {Kelly}}]{vanMeter_2010}
{van Meter}, J.~R., {Miller}, M.~C., {Baker}, J.~G., {Boggs}, W.~D., \&
  {Kelly}, B.~J. 2010, \apj, 719, 1427

\bibitem[{{Voss} \& {Tauris}(2003)}]{Voss2003}
{Voss}, R. \& {Tauris}, T.~M. 2003, \mnras, 342, 1169

\end{thebibliography}

\appendix

\section{Numerical relativity fitting formulae}\label{sec.nrform}

Fitting formulae based on NR are used to calculate the GR merger recoil kick 
and final spin of the merged BH, as discussed in Sect.~\ref{sec.calc}.
The recoil kick, $\vec\vk$, is computed using the fitting formulae
as given in \citet{vanMeter_2010}. These formulae are based on
NR calculations of BH--BH mergers \citep{Baker_2008,vanMeter_2010}
where the component BH spins are generally inclined with respect to the 
orbital axis, resulting in spin-orbit precession (\ie,
precession of the BHs' spin vectors and the orbital axis around the total
angular momentum vector of the BH--BH system) of the merging BH--BH.

The GR merger recoil kick is given by
\beq
\vec\vk = (\vperpm + \vperps\cos\xi)\vec e_1 + \vperps\sin\xi\vec e_2
          + \vpar\vec e_3.
\label{eq:vktot}
\eeq
Here, $\vperpm$ is the mass-ratio-determined recoil,
$\vperps$ is the in-orbital-plane spin-determined recoil,
$\vpar$ is the off-plane (parallel to the orbital axis) spin-determined recoil,
$\vec e_1$ and $\vec e_3$ are the unit vectors along the directions
of the BHs' separation and the orbital axis just before the merger,
respectively, and $\vec e_2 \equiv \vec e_1 \times \vec e_3$.
$\xi=215\degree$ is a constant as obtained in the NR fitting by \citet{Baker_2008}.

The expressions for the in-plane and off-plane recoils are to some
extent motivated by post-Newtonian treatments
\citep{Fitchett_1983,Kidder_1995,Favata_2004,Blanchet_2005,Baker_2007,Gonzalez_2007,Campanelli_2007}
and are given by
\beq
\vperpm = A\eta^2\sqrt{1-4\eta}(1+B\eta), 
\label{eq:vperpm}
\eeq

\beq
\vperps = H\frac{\eta^2}{(1+q)}(a_2\cos\theta_2 - q a_1\cos\theta_1),
\label{eq:vperps}
\eeq

\begin{align}
\vpar & = \frac{K_2\eta^2+K_3\eta^3}{q + 1}\times \nonumber \\
      & \left[q a_1\sin\theta_1\cos(\phi_1-\Phi_1) - a_2\sin\theta_2\cos(\phi_2-\Phi_2)\right] \nonumber  \\
      & + \frac{K_s(q-1)\eta^2}{(q+1)^3}\times  \nonumber \\
      & \left[q^2a_1\sin\theta_1\cos(\phi_1-\Phi_1) + a_2\sin\theta_2\cos(\phi_2-\Phi_2)\right] 
\label{eq:vpar}
\end{align}

In Eqs.~\ref{eq:vperpm}, \ref{eq:vperps}, and \ref{eq:vpar}, $A$, $B$, $H$, $K_2$, $K_3$, and $K_s$
are constants as obtained by NR fitting, $q\equiv m_1/m_2$ ($m_1\leq m_2$) is the
mass ratio of the merging BHs with dimensionless spin magnitudes ($\aone$, $\atwo$),
$\eta\equiv q/(1+q)^2$,
($\theta_1$, $\theta_2$) are the inclination angles that the BHs' spins project to orbital axis,
($\phi_1$, $\phi_2$) are the azimuth angles that the spins project to the separation axis,
and ($\Phi_1$, $\Phi_2$) are phase angles depending on $q$ and the initial separation of the merging BHs'
\citep{Baker_2008,vanMeter_2010}.
Since, in this work, the BHs' spins are assumed to be uncorrelated and randomly oriented
(\ie, $\theta_1$, $\theta_2$, $\phi_1$, $\phi_2$ are random in the range $0\degree-360\degree$)
over a large number of trials (see Sec.~\ref{sec.calc}), the terms $\cos(\phi_1-\Phi_1)$ and
$\cos(\phi_2-\Phi_2)$ will average out in Eq.~\ref{eq:vpar}. Hence, $\Phi_1=\Phi_2=0$
is assumed, as also done in \citet{Morawski2018}. The fitting constants are
$A=1.35\times10^4\kms$, $B=-1.48$, $H=7540\kms$ \citep{Baker_2008} and
$K_2=32092\kms$, $K_3=108897\kms$, $K_s=15375\kms$ \citep{vanMeter_2010}.

The final spin magnitude of the merged BH, $\af$, and its tilt angle with respect to the
orbital axis (orbital angular momentum) at a large separation of the binary, $\thf$, are obtained
using the NR-fitting formulae of \citet{Rezzolla_2008}. The final spin is given by
\begin{align}
\af & = \frac{1}{(1+q)^2}\left[a_2^2 + a_1^2q^4 + 2a_1a_2q^2\cos\alpha \right.\nonumber\\
    & \left. + 2(a_2\cos\theta_2+a_1q^2\cos\theta_1)lq + l^2q^2 \right]^{1/2},
\label{eq:afin}
\end{align}
with \citep[see also,][]{Morawski2018}
\beq
\cos\alpha \equiv \hat a_1 \cdot \hat a_2
= \cos\theta_2\cos\theta_1 + \sin\theta_2\sin\theta_1\cos(\phi_1-\phi_2),
\label{eq:cosa}
\eeq
where $\vec l$ is a vector which is parallel to the orbital angular momentum
at a large binary separation \citep{Rezzolla_2008} and whose magnitude
is given by
\begin{align}
l & = \frac{s_4}{(1+q^2)^2}\left(a_2^2 + a_1^2q^4 + 2a_2a_1q^2\cos\alpha\right) \nonumber\\
  & + \left(\frac{s_5\eta+t_0+2}{1+q^2}\right)\left(a_2\cos\theta_2+a_1q^2\cos\theta_1\right) \nonumber\\
  & + 2\sqrt{3} + t_2\eta + t_3\eta^2.
\label{eq:lmag}
\end{align}
Here, the fitting constants are $s_4 = -0.129$, $s_5 = -0.384$, $t_0 = -2.686$,
$t_2 = -3.454$, and $t_3=2.353$.

The component of the final spin vector parallel to the direction of the (large-separation)
orbital angular momentum is
\beq
\afpar \equiv \vec\af\cdot\hat l = 
\frac{a_2\cos\theta_2 + a_1q^2\cos\theta_1 + lq}{(1+q)^2}
\label{eq:afinpar}
,\eeq
so that the tilt angle of the final spin is given by
$\cos\thf=\afpar/\af$.

We note that in Eqs.~\ref{eq:afin}, \ref{eq:cosa}, \ref{eq:lmag}, and \ref{eq:afinpar},
the  inclination and azimuthal angles of the BH spins are, $\theta_1$, $\theta_2$, $\phi_1$, and $\phi_2$,
are with respect to the binary's orbit and its axis (\ie, the direction of the orbital angular
momentum) at a sufficiently large binary separation, on which the ansatz for
the final spin is based \citep{Rezzolla_2008}.
On the other hand, the angles in the GR recoil kick formulae, as given by
Eqs.~\ref{eq:vperpm}, \ref{eq:vperps}, and \ref{eq:vpar}, are with respect to
the instantaneous orbital plane just before the merger which would be oriented
differently from the large-separation orbital plane if the merging BH--BH system undergoes spin-orbit precession
(see above). However, since, in this work, the spins of the BHs are assumed to be uncorrelated
and randomly oriented over a large number of trials (Sect.~\ref{sec.calc}), the angles can
in effect be used synonymously in both sets of equations.

Also, all of the above equations are expressed following the convention $m_1\leq m_2$
as often employed in NR and as they are implemented in the subroutines used in
this work. The usual precautionary measures are taken to ensure this
convention in these subroutines. However, in the main text of this paper,
the astrophysical convention of $\mone\geq\mtwo$ is followed. 

\end{document}